%% file: paper2_microflares_xsm.tex
\documentclass{aastex63}
\usepackage{upgreek}
\usepackage{amsmath}

\begin{document}

\title{
Observations of the Quiet Sun During the Deepest Solar Minimum of the Past Century with Chandrayaan-2 XSM -- Sub-A Class Microflares Outside Active Regions}
    
\correspondingauthor{Santosh V. Vadawale}
\email{santoshv@prl.res.in}

\author[0000-0002-2050-0913]{Santosh V. Vadawale}
\affiliation{Physical Research Laboratory, Navrangpura, Ahmedabad, Gujarat-380 009, India }
\author[0000-0003-3431-6110]{N. P. S. Mithun}
\affiliation{Physical Research Laboratory, Navrangpura, Ahmedabad, Gujarat-380 009, India }
\affiliation{Indian Institute of Technology Gandhinagar, Palaj, Gandhinagar, Gujarat-382 355, India}
\author[0000-0002-7020-2826]{Biswajit Mondal}
\affiliation{Physical Research Laboratory, Navrangpura, Ahmedabad, Gujarat-380 009, India }
\affiliation{Indian Institute of Technology Gandhinagar, Palaj, Gandhinagar, Gujarat-382 355, India}
\author[0000-0002-4781-5798]{Aveek Sarkar}
\affiliation{Physical Research Laboratory, Navrangpura, Ahmedabad, Gujarat-380 009, India }
\author[0000-0003-2504-2576]{P. Janardhan}
\affiliation{Physical Research Laboratory, Navrangpura, Ahmedabad, Gujarat-380 009, India }
\author[0000-0001-5042-2170]{Bhuwan Joshi}
\affiliation{Physical Research Laboratory, Navrangpura, Ahmedabad, Gujarat-380 009, India }
\author[0000-0003-1693-453X]{Anil Bhardwaj}
\affiliation{Physical Research Laboratory, Navrangpura, Ahmedabad, Gujarat-380 009, India }
\author{M. Shanmugam}
\affiliation{Physical Research Laboratory, Navrangpura, Ahmedabad, Gujarat-380 009, India }
\author[0000-0002-0929-1401]{Arpit R. Patel}
\affiliation{Physical Research Laboratory, Navrangpura, Ahmedabad, Gujarat-380 009, India }
\author{Hitesh Kumar L. Adalja}
\affiliation{Physical Research Laboratory, Navrangpura, Ahmedabad, Gujarat-380 009, India }
\author[0000-0002-3153-537X]{Shiv Kumar Goyal}
\affiliation{Physical Research Laboratory, Navrangpura, Ahmedabad, Gujarat-380 009, India }
\author{Tinkal Ladiya}
\affiliation{Physical Research Laboratory, Navrangpura, Ahmedabad, Gujarat-380 009, India }
\author{Neeraj Kumar Tiwari}
\affiliation{Physical Research Laboratory, Navrangpura, Ahmedabad, Gujarat-380 009, India }
\author{Nishant Singh}
\affiliation{Physical Research Laboratory, Navrangpura, Ahmedabad, Gujarat-380 009, India }
\author{Sushil Kumar}
\affiliation{Physical Research Laboratory, Navrangpura, Ahmedabad, Gujarat-380 009, India }

\begin{abstract}

Solar flares, with energies ranging over several orders of magnitude, result from impulsive release of energy due to magnetic reconnection in the corona. Barring a handful, almost all microflares observed in X-rays are associated with the solar active regions. Here we present, for the first time, a comprehensive analysis of a large sample of quiet Sun microflares  observed in soft X-rays by the Solar X-ray Monitor (XSM)  on board the Chandrayaan-2 mission during the 2019-20 solar minimum. A total of 98 microflares having peak flux below GOES A-level were observed by the XSM during observations spanning 76 days. By using the derived plasma temperature and emission measure of these events obtained by fitting the XSM spectra along with volume estimates from concurrent imaging observations in EUV with the Solar Dynamics Observatory/Atmospheric Imaging Assembly (SDO-AIA), we estimated their thermal energies to be ranging from $3 \times 10^{26}$ to $6 \times 10^{27}$ erg. We present the frequency distribution of the quiet Sun microflares with energy and discuss the implications of these observations of small scale magnetic reconnection events outside active regions on coronal heating.

\end{abstract}


\section{Introduction} 
\label{sec:intro}

Explaining the million Kelvin hot corona over a relatively cooler ($\sim$6000 K) solar photosphere is one of the long-standing problems of
astrophysics. It is fairly well accepted that the magnetic field plays a crucial role in transporting energy from within
the Sun to its atmosphere. However, there is no conclusive evidence, yet, to explain the mechanism involved in the process. 
Two widely considered mechanisms that can heat the corona are heating by Magnetohydrodynamic (MHD) waves and magnetic
reconnection~\citep{2006SoPh..234...41K}.

The large solar flares, which occur due to magnetic reconnection and release total energy ranging from $10^{30}$ to $10^{33}$ erg, can only account for 20\% of the overall coronal energy requirement \citep{2017PJAB...93...87S}. It is known that the solar flares follow a power law distribution and thus the smaller flares occur more frequently, suggesting that even smaller flares may fulfill the energy demand of coronal heating. Based on the early observations of microflares from the Sun (e.g., \citealp{1984ApJ...283..421L}), \citet{1988ApJ...330..474P} hypothesized that the solar corona is heated by the small scale reconnection events, termed as nanoflares having total energy around $\sim 10^{24}$ erg, occurring all over the Sun. 
Since observations of individual small scale reconnection events have not been possible so far due to technological limitations, one possible way to infer their frequency is by extrapolating the power law of the observable microflare distribution to lower energies.
However, ~\citet{1991SoPh..133..357H} showed that 
for the small scale events to be dominant over the large solar flares,  the power law index should be greater than two.
Thus, it is of great importance to constrain the power law index of the flare frequency distribution from the observations of microflares in order to confirm the nanoflare hypothesis.

Several efforts have been made to obtain the frequency distribution of microflares using observations at different wavelengths. In X-rays, the first statistical study of microflares was carried out using the \textit{Yokoh Soft X-ray Telescope} (SXT), where  an active region was observed for five days. Total energy of the microflares observed during this period was inferred from the temperature and emission measure determined using images from two different filters~\citep{shimizu_95} to obtain the frequency distribution within the energy range of $10^{26}$ to $10^{28}$ erg. The most comprehensive statistical study of microflares was carried out using the \textit{Reuven Ramaty High Energy Solar Spectroscopic Imager} (RHESSI) observatory in hard X-ray. This study, spanning over five years~\citep{2008ApJ...677.1385C, 2008ApJ...677..704H} reported more than 25000 microflares, having energies of the order of $10^{27}$erg or higher. The flare parameters were obtained by modeling the X-ray spectrum. 
It should be noted that both the statistical studies were limited to X-ray microflares occuring within active regions.

On the other hand, the proposed nanoflares responsible for heating the corona are expected to be present everywhere on the solar disk, including the quiet corona outside active regions. There have been reports of observations small burst like events in EUV with SOHO EIT~\citep{1998ApJ...501L.213K,2002ApJ...568..413B} and 
TRACE~\citep{2000ApJ...535.1047A,2000ApJ...529..554P}, with total thermal energy in the range of $10^{24} - 10^{26}$erg and following the expected power law distribution. However, it is not clear whether all these events do reach up to coronal temperature or not, due to the difficulties in discriminating impulsive heating events from other brightenings in 
the EUV wavebands~\citep{2000ApJ...535.1027A,2000ApJ...535.1047A}. While this ambiguity does not arise at X-ray energies where the emission arises from higher temperature plasma, there are very few observations due to the technical limitations involved. There have been only a handful of X-ray microflares detected in the quiet Sun with \textit{Yokoh} SXT~\citep{1997ApJ...488..499K}, 
\textit{SphinX}~\citep{2019SoPh..294..176S}, and \textit{NuSTAR}~\citep{2018ApJ...856L..32K}. 
Hence, no statistical studies of quiet Sun X-ray microflares have been possible so far.

The \textit{Solar X-ray Monitor} (XSM) ~\citep{2014AdSpR..54.2021V,shanmugam20} on board
the Chandrayaan-2 mission observed a large number of microflares in the quiet Sun. XSM provides disk integrated X-ray spectral measurements of the Sun with sensitivities down to sub-A class X-ray activity~\citep{2020SoPh..295..139M}. 
It carried out observations of the Sun during 2019-20 solar minimum, which was the deepest in the past 100 years~\citep{2011GeoRL..3820108J, 2015JGRA..120.5306J}, when there were extended periods without any active regions present on the solar disk,  providing a unique opportunity to observe these microflares in the quiet Sun.
Spectroscopic investigation of
non-flaring quiescent corona during this period is discussed in a companion paper~\cite{} (hereafter paper-I). 
Here, we present a detailed study of microflares occurring outside active regions 
as observed by the XSM during this period.
Rest of the paper is organized as follows:
Section~\ref{sec:obs} describes the observations and analysis. Results are presented
and discussed in Section~\ref{sec:result} followed by a summary in Section~\ref{sec:summary}.

\section{Observation of microflares and analysis}
\label{sec:obs}

\begin{figure*}[h!]
\begin{center}
    \includegraphics[width=0.99\textwidth]{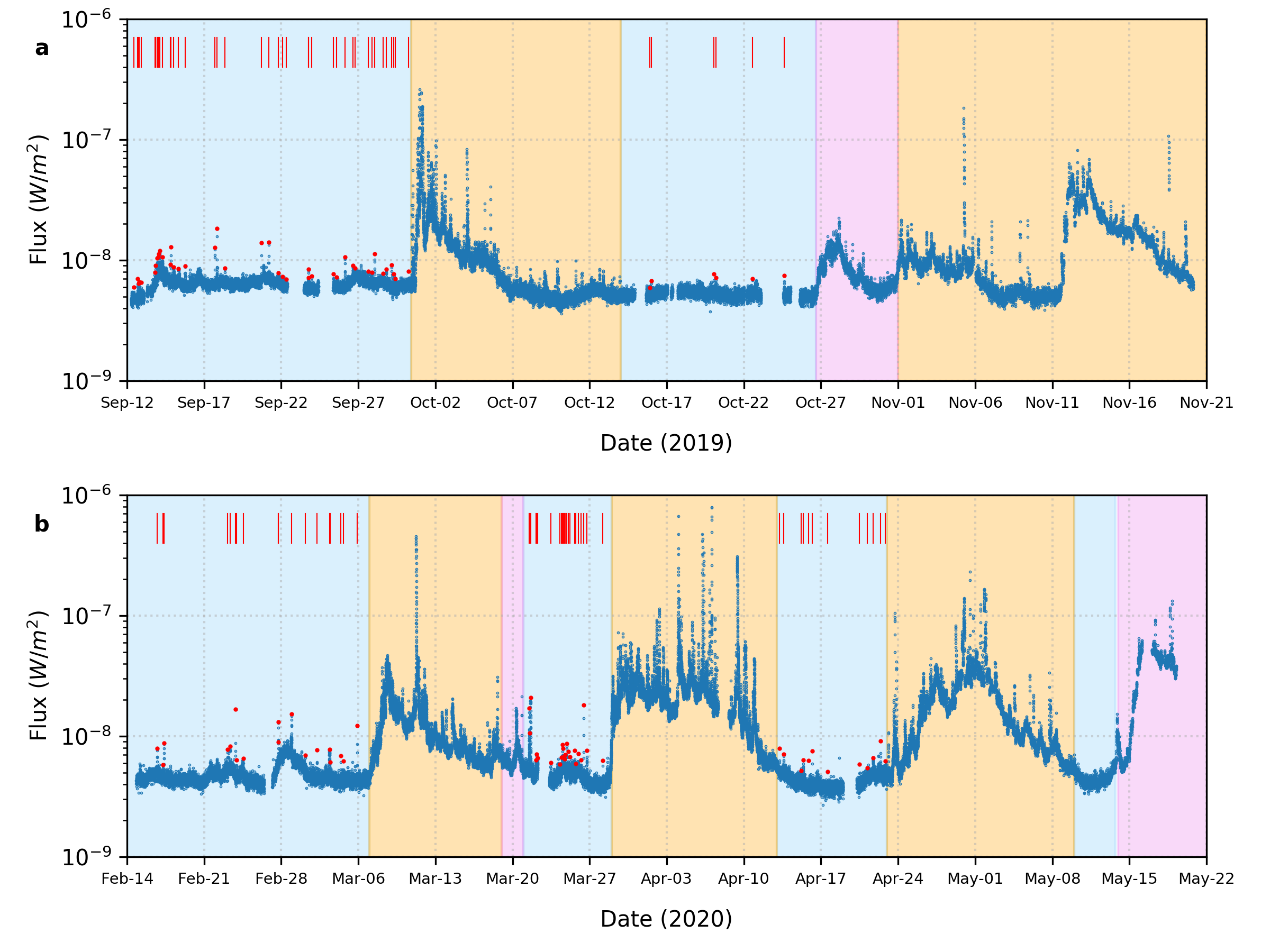}
    \caption{Panels {\bf a} and {\bf b} show the X-ray flux in the 1 -- 15 keV energy range
    with a time cadence of 120 s, as measured by the XSM during two observing seasons.
    Different background shades represent activity levels on the Sun,
    with {\em orange} representing periods when NOAA active regions 
    are present; {\em pink}
    representing periods of enhanced activity visible in both the XSM light curve as well as EUV/X-ray
    images but not classified as AR; and {\em blue} representing periods selected for the
    present study when no major activity was observed on the Sun.
    The microflares detected during the quiet periods are marked with red points, representing their peaks; and red
    vertical bars, representing their time.
    \label{xsmfluxLC}}
\end{center}
\end{figure*}

The Chandrayaan-2 XSM provides soft X-ray spectral measurements of the Sun 
in the energy range of 1--15 keV. 
Here we focus on microflares observed during the periods of extremely low solar activity spanning 76 days, 
identified in Paper-I and shown 
with blue background in Figure~\ref{xsmfluxLC}.

For the selected days, we generated XSM light curves in counts for the 
energy band 1--5 keV with a time bin of two minutes. 
No solar X-ray flux was observed beyond 5 keV due to very low activity.
We also generated light curves in the energy range 
1.5--5 keV as microflares are expected to have a harder spectrum and thus 
easier to detect in the higher energy light curve.
Flare like events were identified from these two light curves by using a 
semi-automated technique of peak identification followed by visual inspection.
For each identified event, the start and end times were obtained from the light curves
and only events having total counts 5$\sigma$ above the average
pre-flare count rate, in either of the light curves, were 
selected. A total of 98 microflares were identified and the peak times these microflares are
marked in Figure~\ref{xsmfluxLC}.

\subsection{EUV counter parts of microflares}

As the spatial location of the microflares are not available from the disk-integrated XSM observations, imaging observations in other wavelengths were used to obtain the same.
For the microflares detected by the XSM, we examined concurrent EUV images of the Sun
obtained with the 
\textit{Solar Dynamics Observatory/Atmospheric Imaging Assembly} (SDO/AIA)~\citep{2012SoPh..275...17L} 
to search for their EUV counter parts. AIA images in the shortest wavelength band
at 94 $\rm{\AA}$ were used for this purpose. Synoptic AIA 94 $\rm{\AA}$ images (Level-1.5) with a cadence of two
minutes available at the Joint Science Operations Center (JSOC) were obtained for the duration
around each microflare. Images were examined to identify any transient events during the 
period of each XSM flare, and they were confirmed by generating light curve for the 
region around the location of the event in AIA and comparing it with the 
XSM light curve. EUV counterparts could be identified for 74 of the 98 microflares 
detected by XSM. 

In order to estimate the volume of flaring region for the flares with identified EUV 
counterparts, we generated maps of Fe XVIII 
emission that trace the plasma having temperature in the range of 3-6 MK, following 
the approach given by \cite{2013A&A...558A..73D}. Fe XVIII maps 
of a 5' $\times$ 5' region surrounding the microflare location
were generated using the AIA images at $94 \rm{\AA}, 211 \rm{\AA},$ and $171 \rm{\AA}$ at the peak of the flares using the equation 
\begin{equation}
I(Fe XVIII) = I(94 \rm{\AA}) - I(211 \rm{\AA})/120 - I(171 \rm{\AA})/450
\end{equation}
From the Fe XVIII images generated in this manner, the pixel having the highest intensity in the flaring region was identified and, those pixels having intensity more than 10\% of the peak intensity were considered to be part of the flare event. The area of the flaring region was calculated from the number of pixels thus identified, and the volume ($V$) of the flaring plasma was estimated as $V = A^{3/2}$.

\subsection{X-ray images and photospheric magnetograms}

In order to examine the X-ray activity and photospheric magnetic field structure associated 
with the microflares, we utilized X-ray images from the \textit{Hinode X-ray Telescope} 
(XRT)~\citep{2007SoPh..243...63G} and  magnetograms from the SDO \textit{Helioseismic and Magnetic Imager} 
(HMI)~\citep{2012SoPh..275..207S}. From the available full disk images obtained with Hinode XRT 
and hourly synoptic magnetograms with SDO HMI, observations before and after the peak time of 
all microflares with EUV counter parts were selected.
For XRT we chose images with the Be-thin filter as its efficiency curve in the low energy range 
is very similar to that of the XSM and thus these images are expected to represent the 
spatial distribution of X-ray emission observed by the XSM. 
Cutouts of the XRT images and HMI magnetograms for the flare location
were generated for each case after taking into account the solar rotation.
It may be noted that XRT images are not available at uniform intervals and in some cases the differences
between the X-ray image time and the flare time are more than a day. It may also be
noted that the photospheric magnetograms for near limb events may not be very reliable.

\subsection{X-ray spectroscopy of microflares}

We analyzed the XSM soft X-ray spectra of microflares to determine 
the flare temperature and emission measure. 
For each microflare, an interval around the peak that covers 50\% 
of the flare duration was determined. Integrated spectra for the selected duration was generated using the XSM Data Analysis Software (XSMDAS). Spectral fitting 
was carried out using XSPEC package~\citep{arnaud96} with an isothermal plasma emission model 
based on CHIANTI atomic database version 9.0.1  ~\citep{1997A&AS..125..149D,2019ApJS..241...22D}, 
similar to the quiet Sun spectral analysis discussed in paper-I.  
The integrated spectrum of a microflare consists of emission from the flare and emission from the pre-flare quiescent Sun.
Usually, spectral analysis is carried out by subtracting the pre-flare spectrum from that of the flare. 
However, for XSM, this may not be always appropriate as the angle subtended by the Sun with the XSM boresight varies over an orbit and thus the effective area. Therefore, instead of subtracting the quiescent emission, we obtained the quiet Sun spectral model of the respective day by fitting the  observed quiet Sun spectrum for the periods excluding the flare duration with sufficient margin, as shown in Figure 2 of paper-I. 
It may be noted that the overall quiet Sun spectrum did not change significantly over a day. 
We then used a two-component model for fitting the microflare spectra, where one component is fixed to the quiet Sun spectral model and the other one belongs to the flare. 
For the second component corresponding to the flare emission, only
the temperature and emission measure were left as free parameters in fitting 
while the abundances were frozen to the quiet Sun values as it is not possible 
to constrain elemental abundances due to low statistics of the microflare spectra. 
We note that the statistical uncertainties on the quiet Sun spectral 
model parameters are much smaller. We have verified that they have minimal impact on the estimated 
microflare parameters and their uncertainties, obtained by fixing the quiet Sun component at its 
best fit values.     

\section{Results and Discussion}
\label{sec:result}

The XSM light curves presented in Figure~\ref{xsmfluxLC} shows several 
flaring events throughout the observation. 
During the 76 days identified when solar active regions were absent,
we detected 98 microflares from XSM observations having effective exposure 
of 53.3 days, resulting in an average number of events to be $\sim 1.84~\rm{{day}^{-1}}$.
Mean event rates vary from $\sim 0.75~\rm{{day}^{-1}}$ to $\sim 3.4~\rm{{day}^{-1}}$ 
for individual epochs of quiet Sun observations.
These events are designated with IDs corresponding to the peak time, 
following the standard convention~\citep{2010SoPh..263....1L} and are listed 
in the Supplemetary Information.
While a small number of such X-ray microflares occurring outside ARs have
been reported earlier with Yokoh (4 microflares,~\citealp{1997ApJ...488..499K}), 
SphinX (16 microflares,~\citealp{2019SoPh..294..176S}), as well as recently with 
NuSTAR (3 microflares,~\citealp{2018ApJ...856L..32K}), this
is the first observation in X-ray wavelengths of such a large number of 
microflares occurring outside active regions, with the observations spanning a few months. 
This  demonstrates that microflares are not confined only to the active 
regions and supports the hypothesis on presence of small scale impulsive events 
everywhere on the solar corona.

\begin{figure*}[h!]
\begin{center}
    \includegraphics[width=0.99\textwidth]{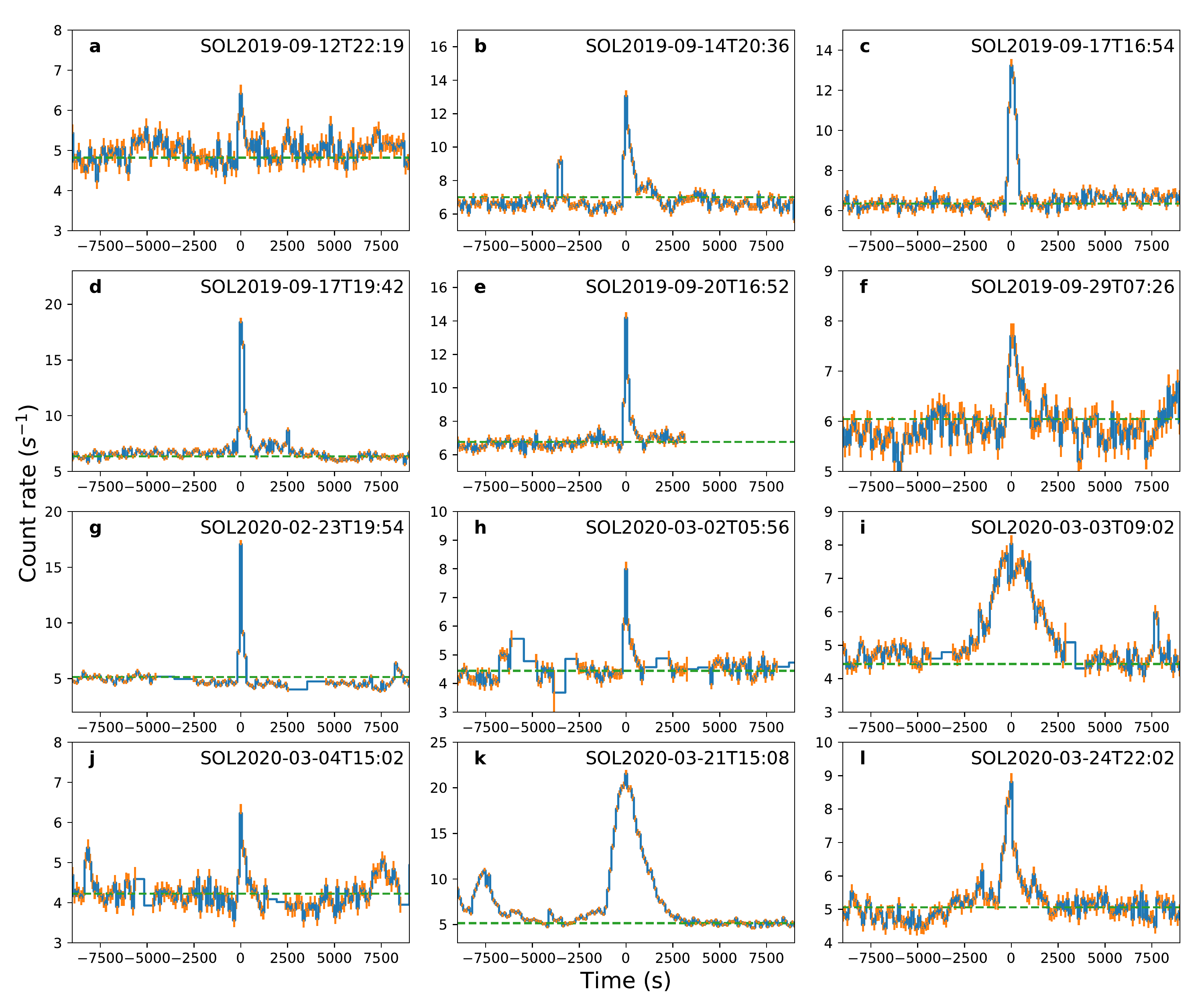}
    \caption{X-ray light curves of a representative set of microflares observed by the XSM in the 
    energy range of 1--5 keV. Flare IDs are shown in the respective panels. Error bars correspond to one sigma uncertainties. Green dashed lines show the mean count rate for the duration considered for non-flaring quiescent emission.}
    \label{xsmFlareLc}
\end{center}
\end{figure*}

X-ray light curves for a representative set of 
these microflares are plotted in Figure~\ref{xsmFlareLc}.
Most of these microflares display a normal flare like behavior with a 
fast rise and slow decay, suggesting an impulsive energy release.
However, there are a few microflares that do not follow this behavior possibly
due to the blending of multiple microflares or due to these having an 
intrinsically different origin.

\subsection{Microflare location}

\begin{figure*}[h!]
\begin{center}
    \includegraphics[width=0.99\textwidth]{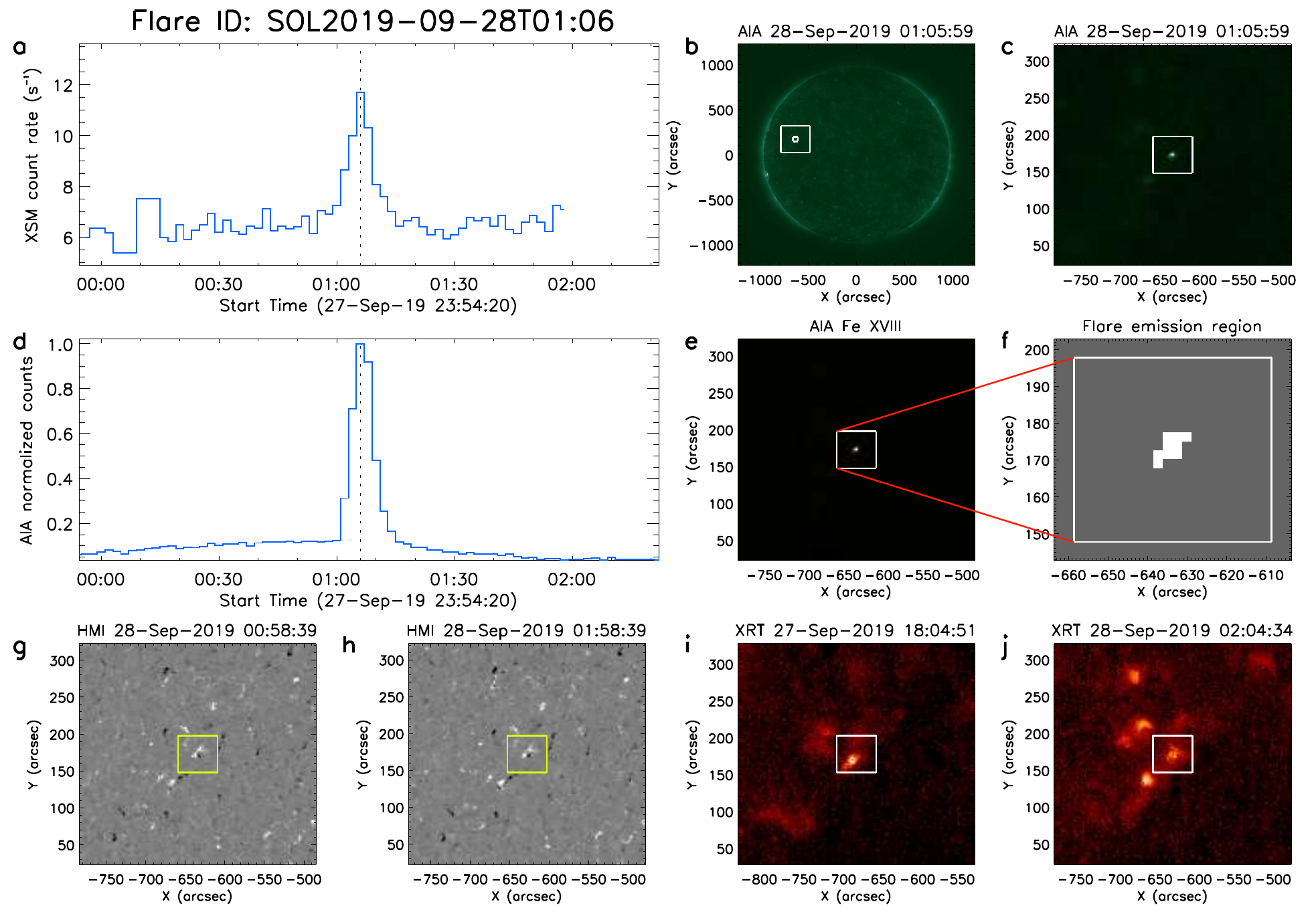}
    \caption{Identification of the location of one of the XSM microflares with 
    the 1--5 keV light curve shown in panel {\bf a}. Flare location is marked on 
    the SDO AIA 94$~\rm{\AA}$ full disk image in {\bf b} and a 5' $\times$ 5' 
    cutout is shown in {\bf c}. 
    Panel {\bf d} shows AIA 94$~\rm{\AA}$ light curve for the flaring pixels 
    as shown in panel {\bf f}. 
    FeXVIII image of the flare location is shown in panel {\bf e} and the map of pixels of 
    the flaring plasma based on FeXVIII emission is given in {\bf f}.         
    Available synoptic HMI magnetograms and XRT Be-thin images nearest to the flare peak time are shown in panels 
    {\bf g-j}. The complete figure set (for all 98 microflares) is available in the online journal.
    }
    \label{xsmFlareCounterPart}
\end{center}
\end{figure*}

We identified the location on the solar disk for 74 of the XSM observed microflares using AIA 
94 $\rm{\AA}$ images during the flare duration. Association of the X-ray events with the EUV 
counterpart was confirmed by comparison of the light curves. 
Figure~\ref{xsmFlareCounterPart} shows
an example of the EUV counterpart identification for one microflare where the light curve for XSM 
is plotted in ~\ref{xsmFlareCounterPart}a and the AIA light curve for the identified region 
(marked in images in ~\ref{xsmFlareCounterPart}b and ~\ref{xsmFlareCounterPart}c) is shown 
in ~\ref{xsmFlareCounterPart}d. Similar figures for all microflares are available in the  
Supplementary Information. 
From the Fe XVIII images obtained from three AIA bands, we find 
that the flaring region is often very small, limited to few pixels in the images. However, 
there are also some cases where complex structures are resolved in the images. For uniformity, 
the emission volume is estimated based on number of identified pixels for all microflares. 

Further, for the microflares with identified location, we examined the photospheric magnetic
fields and X-ray activity before and after the microflare using synoptic data from HMI and XRT as 
shown in Figure  ~\ref{xsmFlareCounterPart}g- ~\ref{xsmFlareCounterPart}j for one event. 
We find that, wherever reliable magnetograms were available, the microflares were associated
with magnetic bipolar regions having weaker field strengths compared to active regions.
Distinct association of microflares with bipolar regions provide an indication for
magnetic reconnection in coronal loops, as in standard flare model, but at much smaller scales.
We also find that most of these microflares are associated with X-ray Bright Points 
(XBPs, \citealp{1974ApJ...189L..93G}) seen in XRT images. 
However, there are some microflares for which XBPs are observed before the
event but not after (see flare id SOL2019-09-17T16:54 in
supplementary figure) and vice-versa (e.g. flare id SOL2019-09-13T23:06). These
observations can play a vital role in understanding the formation and
evolution of the XBPs\citep{1994ApJ...427..459P,2019LRSP...16....2M}.
More interestingly, we also find a few flares which do not appear to be
associated with any XBP (e.g. flare id SOL2020-04-20T12:11); however, it is
difficult to draw any specific conclusion given the
large gap and non-uniform sampling between successive X-ray images. 

\subsection{Temperature and Emission Measure}

\begin{figure}
\begin{center}
    \includegraphics[width=1.0\columnwidth]{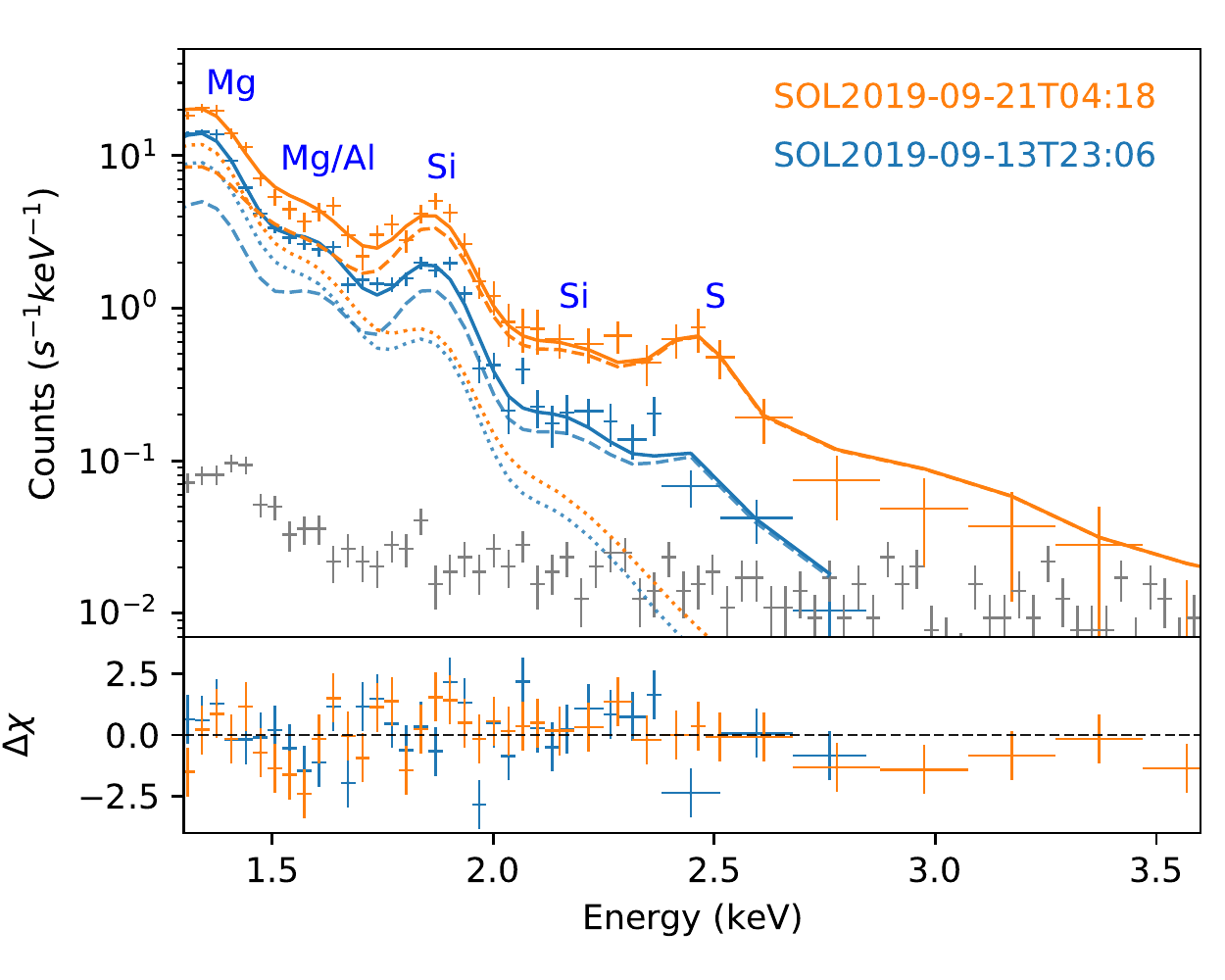}
    \caption{The XSM soft X-ray spectra for two representative microflares are shown with the fitted model.
 The best fit models for microflares, shown with solid lines, consists of two components:
 one corresponding to the background quiet Sun emission (dotted lines) and the other corresponding to the
emission from the flaring plasma (dashed lines). 
Residuals are shown in the bottom panel. The gray color points represent 
the non-solar X-ray background spectrum.
    \label{xsmSpecFit}}
\end{center}
\end{figure}

By modeling the soft X-ray spectra of the microflares observed with the XSM, we obtain 
their temperature and emission measure (EM). Figure~\ref{xsmSpecFit} shows the XSM spectra 
for two representative microflares along with the best fit models. It may be noted
that the energy range for fitting is restricted  typically up to 3-4 keV, where the solar spectrum dominates 
over the non-solar background. 
As seen in the figure, the spectra are well fitted with 
the two component model. Similar fits were obtained 
for other events as well. We could obtain measurements of temperature and EM with robust 
uncertainties for 86 of the 98 microflares, while the parameters could not be constrained 
for the remaining due to low statistics and hence no measurements are reported for them.
Estimated spectral parameters along with other details for all the microflares are
tabulated in the appendix.

\begin{figure*}[h!]
\begin{center}
    \includegraphics[width=0.99\textwidth]{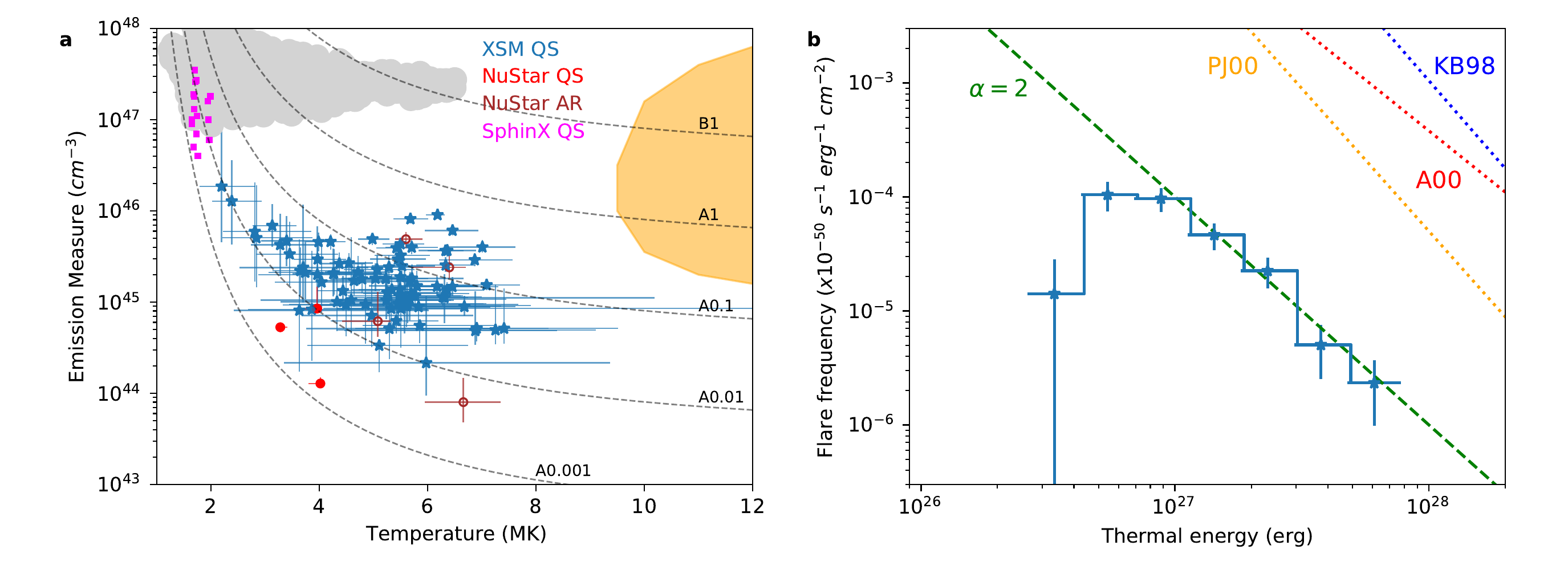}
    \caption{
    Panel {\bf a} shows temperature and EM for 86 of the 98 quiet Sun (QS) microflares observed by
    the XSM with blue star symbols. The error bars represent 1$\sigma$ uncertainties.
    Magneta squares and red filled circles correspond to QS microflares observed by SphinX~\citep{2019SoPh..294..176S} 
    and NuSTAR~\citep{2018ApJ...856L..32K}, respectively.
    Parameter space 
    of active region (AR) events observed by RHESSI~\citep{2008ApJ...677..704H} and SphinX~\citep{2017SoPh..292...77G} 
    are shown with orange and gray shades, 
    respectively.
    The brown open circles represent the four AR microflares reported by 
    NuSTAR~\citep{wright_17,2017ApJ...845..122G,2019ApJ...881..109H,2020ApJ...893L..40C}.
    Dashed lines represent the isoflux curves corresponding to GOES/XRS $1-8~\rm{\AA}$ X-ray flux levels from
    A0.001 ($10^{-11}~\rm{Wm^{-2}}$) to B1 ($10^{-7}~\rm{Wm^{-2}}$).
    Panel {\bf b} shows the frequency distribution of thermal energies of 63
    microflares for which the measurements are available. 
    The green dashed line corresponds to a power law index ($\alpha$) of two shown for 
    comparison. The dotted lines correspond to power laws reported from quiet Sun EUV observations 
    by \citet{1998ApJ...501L.213K}[KB98], \citet{2000ApJ...535.1047A} [A00], and \citet{2000ApJ...529..554P} [PJ00] as shown in Figure 10 of \citet{2000ApJ...535.1047A}.
    \label{xsmFlarePar}}
\end{center}
\end{figure*}

Figure~\ref{xsmFlarePar}a shows EM vs temperature of the XSM microflares 
along with the reported measurements for microflares  observed by other 
instruments. Isoflux curves corresponding to 1--8 $\rm{\AA}$ flux are 
shown with dashed lines in the figure and it can be seen from the figure 
that XSM detects microflares above peak a flux of 
$\sim 10^{-10} ~\rm{W m^{-2}}$. 
From the figure, it can be seen that 
the quiet Sun microflares observed by the XSM have lower temperature, 
EM, and thus flux compared to the sample of active region microflares 
observed with RHESSI~\citep{2008ApJ...677..704H}.
However, the XSM microflare parameters are similar
to those observed by NuSTAR, even though four of the seven NuSTAR
microflares have been observed within active regions.
This suggests that the basic physical processes in the non-AR
microflares observed by XSM are similar to those in the active
region microflares and may differ only in terms of the scale.

Two of the quiet Sun microflares observed by NuSTAR~\citep{2018ApJ...856L..32K} 
are weaker than all the events observed by the XSM, which is expected given 
the sensitivity of NuSTAR. We also note that the 16 microflares observed by 
SphinX show a systematically lower temperature and higher EM compared to 
the microflares observed by XSM as well as NuSTAR. One possible reason for 
this observation may be the difference in analysis procedure followed for SphinX data where 
the flare spectra were fitted with a single temperature model 
without subtracting or separately modelling the pre-flare emission~\citep{2019SoPh..294..176S}.
While the microflares observed by SphinX
and XSM occupy slightly different regions in the temperature-EM parameter space, the fact that they fall 
along the same isoflux line shows that they should be of similar nature.

\subsection{Microflare thermal energy distribution}

Using the temperature ($T$) and emission measure ($EM$) obtained from X-ray spectroscopy and the
plasma volume ($V$) obtained from the AIA images, we estimated the thermal energy associated
with each microflare~\citep{2008ApJ...677..704H} as

\begin{equation}
E_{th} \sim 3~k_{\rm{B}}T~\sqrt{EM\times V}
\end{equation}

\noindent where $k_{\rm{B}}$ is the Boltzmann constant. 
This assumes a filling factor of unity, which is an upper limit. Further, 
as the volume is estimated from lower energy emission, it also can be considered as an
upper limit to the actual volume and thus the energy estimates represent upper limits. 
Of the total 98 microflares, thermal energy
could be estimated for 63 flares for which measurements of all three parameters
are available (see appendix).
Errors on thermal energy were computed by propagating the errors on
temperature and emission measure.
We find that the thermal energies of XSM microflares ranges from $3\times10^{26} - 7\times10^{27} \rm{erg}$. 
A histogram of thermal energies of the microflares was then generated and is normalized with the
total exposure time and area, which is shown in Figure~\ref{xsmFlarePar}b. 
Error bars on the histogram are computed considering counting statistics. 
A power law of index two is overplotted in the figure with dashed line for comparison and 
it can be seen that frequency distribution agrees with the power law. 
The Departure from power law seen at the lower energy end is expected from the fact that
fainter events are less likely to be detected against the background quiet Sun emission. A similar
trend has been seen in the distribution of active region flares from previous studies~\citep{2011SSRv..159..263H}.

Even though the present observations provide the largest sample of microflares in the quiet Sun, the total number is not sufficient for quantitative estimation of the power law index. 
However, Figure~\ref{xsmFlarePar}b shows with certainty that the observed histogram is significantly different from the power law derived from previous quiet Sun EUV nanoflare observations~\citep{1998ApJ...501L.213K,2000ApJ...535.1047A,2000ApJ...529..554P} 
extrapolated to higher energies. 
It should be noted that two of the flare frequency histograms in EUV are obtained during solar minimum in 1998-99 and thus the difference may not be due to solar activity. This suggests that the difference may be either due to a change in power law index over the typical nanoflare to microflare energies or the EUV and X-ray events are from two different populations. %
In any case, our observations, being the first statistical survey of X-ray 
microflares outside active regions, are expected to provide crucial inputs 
to understand the role of small scale impulsive events in coronal heating.

Another point to note is that the present work, like most others, considered 
only the energy associated with the thermal emission, The total 
energy may include other components such as non-thermal accelerated particles 
and waves and it is important to understand the energy partition into 
these~\citep{2017LRSP...14....2B}. 
With sensitive X-ray spectroscopic observations, it has been possible to 
disentangle thermal and non-thermal components in A class microflare~\citep{2020ApJ...891L..34G}.
Broad-band X-ray observations of small scale events using the XSM along with 
hard X-ray instruments such as Solar Orbiter STIX~\citep{2020A&A...642A..15K} and NuSTAR 
will help in improving the understanding of thermal and non-thermal energy associated 
with such events, so that better estimates of total energy can be obtained.

\section{Summary}
\label{sec:summary}

We presented the largest sample of X-ray microflares in the quiet
Sun using disk-integrated observations with the Chandrayaan-2 XSM. 
From XSM observations during extended periods without active regions 
on the solar disk, 98 sub-A class microflares were detected. 
We find that most of the microflares are impulsive in nature and in all cases where the microflare location is identified, they are associated with magnetic bipolar regions.
Using the results of spectroscopic analysis of the XSM data and analysis of EUV images, we estimated that the thermal energies of these microflares range from 
$\sim3\times10^{26}$ -- $6\times10^{27}$ erg and the flare frequency 
distribution follows a power law. 
The present observations provide 
a stringent limit on the average number of microflares 
having flux above $\sim 10^{-10} ~\rm{W m^{-2}}$ (thermal energies above $\sim 4-7\times 10^{26}$erg) ocurring in the quiet Sun. During the period of observed extremely quiet solar corona, having no active region on the disk, the flare rate is found to be $\sim 1.84~\rm{{day}^{-1}}$, while the mean values for each epoch vary from $\sim 0.75~\rm{{day}^{-1}}$ to $\sim 3.4~\rm{{day}^{-1}}$. 
While the statistics may not be sufficient 
to constrain the index of the power law very well, these observations provide 
strong support to the hypothesis of occurrence of small scale impulsive heating 
events everywhere on the solar disk,  which could contribute to the heating of the corona.

\acknowledgments
{
XSM was designed and developed by the Physical Research Laboratory (PRL), Ahmedabad
with support from the  Space Application Centre (SAC), Ahmedabad,
the U. R. Rao Satellite Centre (URSC), Bengaluru, and the Laboratory for Electro-Optics
Systems (LEOS), Bengaluru. 
We thank various facilities and the technical teams of all
the above centers and Chandrayaan-2 project, mission operations, and ground segment teams
for their support.
The Chandrayaan-2 mission is funded and managed by the Indian Space Research Organisation (ISRO).
Research at PRL is supported by the Department of Space, Govt. of India.
This work made use of synoptic images obtained with AIA and HMI instruments on board SDO as well as
that from XRT on board Hinode.
We gratefully acknowledge G. Del-Zanna, H. E. Mason, and U. Mitra Kraev for their help with the use of CHIANTI,
EUV analysis, as well as very helpful discussions, facilitated through Royal Society Grant No. IES{\textbackslash}R2{\textbackslash}170199.
}
\vspace{5mm}
\facilities{Chandrayaan-2(XSM), SDO(AIA, HMI), Hinode(XRT)}


\software{XSMDAS~\citep{mithun20_soft}, XSPEC~\citep{arnaud96}, SolarSoft}

\appendix

\section{Microflare Parameters}

{\def\arraystretch{1.2}

\begin{longrotatetable}
 \begin{deluxetable*}{c c c c c c c c c}
 \tablecaption{Micoflares observed by XSM in the quiet Sun period (blue shaded region in Figure 1). \\
 $^a$Flare ID correspond to the time at
 the peak of the flare in the format SOL$yyyy$-$mm$-$dd$T$hh$:$mm$.\\
 $^b$Peak count rate and peak flux include the pre-flare emission rate/flux. \\
 $^c$Temperature, volume EM, and flare flux are estimated for the flaring plasma alone.
 Pre-flare emission is modelled as a separate
 component as discussed in the text.\\
 $^d$Number of pixels in AIA FeXVIII image at the peak of the flare providing the area ($A$) of the flaring region.
 Each pixel correspond to area of 2.4" x 2.4". Volume ($V$) of the flaring plasma is estimated as $V = A^{3/2}$.\\
 $^e$Thermal energy estimated for flares with temperature, EM, and volume measurement available.\\
 $^*$Flux values are shown in units of $10^{-8}~\mathrm{Wm^{-2}}$ corresponding to A level of GOES/XRS X-ray intensity scale.\\
 }
 \label{xsm_flare_list}
\tablewidth{700pt}
\tabletypesize{\scriptsize}
\tablehead{
 \colhead{Sl No } & \colhead{ Flare ID$^a$ } & \colhead{ Peak rate$^{b}$  } & \colhead{ Peak 1-8$\mathrm{\AA}$ flux$^{b*}$ } & \colhead{ Temperature$^c$ } & \colhead{ Volume EM$^c$ } & \colhead{ Flare 1-8$\mathrm{\AA}$ flux$^{c*}$ } & \colhead{ No. of pixels$^d$ } & \colhead{ $\mathrm{E_{thermal}}^e$} \\
\colhead{   } & \colhead{  } & \colhead{ ($\mathrm{s^{-1}}$)} & \colhead{ (x$10^{-8}~\mathrm{Wm^{-2}}$) } & \colhead{ (MK) } & \colhead{ (x$10^{46}~\mathrm{{cm}^{-3}}$) } & \colhead{(x$10^{-8}~\mathrm{Wm^{-2}}$) } & \colhead{ AIA FeXVIII } & \colhead{ (erg)} \\
}
\startdata
\input{xsm_flare_table_data.tex}
\enddata
\end{deluxetable*}
\end{longrotatetable}
}

\bibliographystyle{aasjournal}

\end{document}

%% file: xsm_flare_table_data.tex
      1&SOL2019-09-12T11:01& 6.7& 0.09&$ 3.6\substack{+ 2.4\\-1.2}$&$ 0.08\substack{+ 0.41\\-0.06}$& 0.007&          32&$  1.3\substack{+  3.4\\ -0.7}\times 10^{27}$\\
      2&SOL2019-09-12T16:07& 8.4& 0.13&$ 5.8\substack{+ 1.3\\-0.8}$&$ 0.09\substack{+ 0.04\\-0.03}$&  0.039&- & - \\
      3&SOL2019-09-12T18:33& 7.9& 0.11&$ 7.3\substack{+ 1.9\\-2.9}$&$ 0.05\substack{+ 0.11\\-0.02}$& 0.037&          29&$  1.9\substack{+  2.1\\ -0.8}\times 10^{27}$\\
      4&SOL2019-09-12T22:19& 7.6& 0.13& - & - & - &          10 & - \\
      5&SOL2019-09-13T19:16& 9.3& 0.14&$ 3.4\substack{+ 0.6\\-0.5}$&$ 0.47\substack{+ 0.41\\-0.22}$& 0.029&          11&$  1.3\substack{+  0.6\\ -0.4}\times 10^{27}$\\
      6&SOL2019-09-13T20:38&10.6& 0.18&$ 5.0\substack{+ 0.7\\-0.6}$&$ 0.18\substack{+ 0.08\\-0.05}$& 0.052&          10&$  1.1\substack{+  0.3\\ -0.2}\times 10^{27}$\\
      7&SOL2019-09-13T23:06&12.0& 0.21&$ 4.0\substack{+ 0.3\\-0.3}$&$ 0.46\substack{+ 0.12\\-0.10}$& 0.057&          39&$  4.0\substack{+  0.6\\ -0.5}\times 10^{27}$\\
      8&SOL2019-09-14T00:42&12.4& 0.20&$ 5.1\substack{+ 0.7\\-0.6}$&$ 0.23\substack{+ 0.10\\-0.06}$&  0.067&- & - \\
      9&SOL2019-09-14T01:48&13.3& 0.24&$ 5.7\substack{+ 0.5\\-0.5}$&$ 0.17\substack{+ 0.05\\-0.03}$& 0.069&          22&$  2.2\substack{+  0.4\\ -0.3}\times 10^{27}$\\
     10&SOL2019-09-14T03:24&14.2& 0.23&$ 4.2\substack{+ 0.3\\-0.2}$&$ 0.46\substack{+ 0.09\\-0.08}$&  0.071&- & - \\
     11&SOL2019-09-14T07:04&12.3& 0.20&$ 4.5\substack{+ 0.4\\-0.4}$&$ 0.27\substack{+ 0.08\\-0.06}$& 0.055&          33&$  3.1\substack{+  0.5\\ -0.4}\times 10^{27}$\\
     12&SOL2019-09-14T19:38&10.4& 0.19&$ 6.9\substack{+ 1.3\\-1.1}$&$ 0.05\substack{+ 0.00\\-0.01}$& 0.034&           5&$  4.9\substack{+  1.0\\ -1.0}\times 10^{26}$\\
     13&SOL2019-09-14T20:36&14.6& 0.32&$ 5.5\substack{+ 0.6\\-0.6}$&$ 0.25\substack{+ 0.07\\-0.05}$& 0.095&           6&$  1.0\substack{+  0.2\\ -0.1}\times 10^{27}$\\
     14&SOL2019-09-15T00:42& 9.7& 0.30&$ 3.6\substack{+ 0.7\\-0.5}$&$ 0.22\substack{+ 0.18\\-0.10}$&  0.019&- & - \\
     15&SOL2019-09-15T08:16& 9.9& 0.16&$ 4.0\substack{+ 0.5\\-0.4}$&$ 0.30\substack{+ 0.15\\-0.10}$&  0.036&- & - \\
     16&SOL2019-09-17T16:54&14.9& 0.30&$ 5.7\substack{+ 0.5\\-0.5}$&$ 0.40\substack{+ 0.09\\-0.06}$& 0.166&           5&$  1.1\substack{+  0.2\\ -0.1}\times 10^{27}$\\
     17&SOL2019-09-17T19:42&20.3& 0.59&$ 7.0\substack{+ 0.6\\-0.7}$&$ 0.40\substack{+ 0.08\\-0.05}$&  0.276&- & - \\
     18&SOL2019-09-18T08:36&10.1& 0.17&$ 5.5\substack{+ 1.0\\-0.8}$&$ 0.14\substack{+ 0.07\\-0.04}$&  0.051&- & - \\
     19&SOL2019-09-20T16:52&15.8& 0.36&$ 6.3\substack{+ 0.5\\-0.5}$&$ 0.25\substack{+ 0.06\\-0.04}$& 0.139&          35&$  4.3\substack{+  0.6\\ -0.5}\times 10^{27}$\\
     20&SOL2019-09-21T04:18&16.1& 0.38&$ 6.4\substack{+ 0.5\\-0.5}$&$ 0.37\substack{+ 0.07\\-0.05}$&  0.203&- & - \\
     21&SOL2019-09-21T19:18& 9.4& 0.11&$ 2.2\substack{+ 0.6\\-0.4}$&$ 1.86\substack{+ 5.86\\-1.40}$& 0.008&           9&$  1.5\substack{+  2.4\\ -0.6}\times 10^{27}$\\
     22&SOL2019-09-22T01:58& 8.8& 0.11& - & - & - &          17 & - \\
     23&SOL2019-09-22T07:32& 8.4& 0.11& - & - & - &           7 & - \\
     24&SOL2019-09-23T18:04& 8.4& 0.13&$ 4.0\substack{+ 1.5\\-1.1}$&$ 0.20\substack{+ 0.48\\-0.12}$&  0.025&- & - \\
     25&SOL2019-09-23T18:32& 9.9& 0.15&$ 6.3\substack{+ 3.9\\-1.4}$&$ 0.11\substack{+ 0.09\\-0.05}$&  0.060&- & - \\
     26&SOL2019-09-23T23:04& 8.4& 0.13&$ 6.9\substack{+ 1.5\\-2.4}$&$ 0.05\substack{+ 0.08\\-0.01}$& 0.032&           6&$  5.5\substack{+  4.8\\ -2.1}\times 10^{26}$\\
     27&SOL2019-09-25T08:50& 8.7& 0.11&$ 5.0\substack{+ 1.9\\-1.3}$&$ 0.07\substack{+ 0.11\\-0.04}$& 0.020&           6&$  4.8\substack{+  4.1\\ -1.8}\times 10^{26}$\\
     28&SOL2019-09-26T03:32&12.3& 0.21&$ 7.1\substack{+ 0.6\\-0.5}$&$ 0.15\substack{+ 0.02\\-0.02}$& 0.109&          12&$  1.7\substack{+  0.2\\ -0.2}\times 10^{27}$\\
     29&SOL2019-09-26T15:50&10.8& 0.17&$ 4.4\substack{+ 0.5\\-0.4}$&$ 0.26\substack{+ 0.09\\-0.07}$&  0.047&- & - \\
     30&SOL2019-09-26T18:34&10.3& 0.12&$ 3.1\substack{+ 0.4\\-0.4}$&$ 0.69\substack{+ 0.50\\-0.28}$&  0.029&- & - \\
     31&SOL2019-09-27T15:32& 9.5& 0.14&$ 4.5\substack{+ 1.5\\-1.2}$&$ 0.09\substack{+ 0.17\\-0.05}$&  0.018&- & - \\
     32&SOL2019-09-27T21:00& 9.5& 0.10&$ 5.1\substack{+ 1.6\\-1.3}$&$ 0.03\substack{+ 0.00\\-0.02}$&  0.010&- & - \\
     33&SOL2019-09-28T01:06&13.2& 0.24&$ 5.5\substack{+ 0.5\\-0.5}$&$ 0.30\substack{+ 0.08\\-0.06}$& 0.109&           9&$  1.5\substack{+  0.2\\ -0.2}\times 10^{27}$\\
     34&SOL2019-09-28T14:16& 9.2& 0.11&$ 2.4\substack{+ 0.6\\-0.4}$&$ 1.28\substack{+ 2.32\\-0.85}$& 0.011&          17&$  2.1\substack{+  2.0\\ -0.8}\times 10^{27}$\\
     35&SOL2019-09-28T19:42& 9.9& 0.13&$ 4.8\substack{+ 0.7\\-0.6}$&$ 0.18\substack{+ 0.08\\-0.06}$& 0.044&          16&$  1.5\substack{+  0.4\\ -0.3}\times 10^{27}$\\
     36&SOL2019-09-29T04:04&10.4& 0.19&$ 5.6\substack{+ 2.3\\-1.7}$&$ 0.09\substack{+ 0.17\\-0.05}$& 0.036&           8&$  7.6\substack{+  7.8\\ -3.0}\times 10^{26}$\\
     37&SOL2019-09-29T07:26& 9.1& 0.13&$ 7.4\substack{+ 2.1\\-2.7}$&$ 0.05\substack{+ 0.09\\-0.02}$&  0.040&- & - \\
     38&SOL2019-09-29T09:54& 8.2& 0.13& - & - & - & - & - \\
     39&SOL2019-09-30T06:22& 9.4& 0.17&$ 4.9\substack{+ 2.8\\-1.4}$&$ 0.09\substack{+ 0.18\\-0.06}$& 0.024&           8&$  6.7\substack{+  7.6\\ -2.9}\times 10^{26}$\\
     40&SOL2019-10-15T21:25& 7.2& 0.08& - & - & - &          13 & - \\
     41&SOL2019-10-15T23:47& 8.0& 0.11& - & - & - &           8 & - \\
     42&SOL2019-10-20T01:34& 8.8& 0.13&$ 3.7\substack{+ 1.2\\-1.2}$&$ 0.24\substack{+ 0.93\\-0.14}$& 0.022&           6&$  6.5\substack{+ 12.8\\ -2.8}\times 10^{26}$\\
     43&SOL2019-10-20T04:38& 8.1& 0.17&$ 4.3\substack{+ 1.3\\-1.0}$&$ 0.10\substack{+ 0.15\\-0.05}$& 0.017&          10&$  7.3\substack{+  5.8\\ -2.6}\times 10^{26}$\\
     44&SOL2019-10-22T13:18& 8.3& 0.12&$ 5.7\substack{+ 1.1\\-0.9}$&$ 0.12\substack{+ 0.06\\-0.04}$& 0.051&           8&$  9.0\substack{+  2.8\\ -1.9}\times 10^{26}$\\
     45&SOL2019-10-24T14:32& 8.4& 0.20& - & - & - &           6 & - \\
     46&SOL2020-02-16T18:00& 9.2& 0.18&$ 6.5\substack{+ 0.9\\-0.7}$&$ 0.15\substack{+ 0.04\\-0.03}$& 0.084&           7&$  1.0\substack{+  0.2\\ -0.1}\times 10^{27}$\\
     47&SOL2020-02-17T06:30& 6.9& 0.10&$ 4.0\substack{+ 1.2\\-0.9}$&$ 0.17\substack{+ 0.23\\-0.09}$& 0.022&          29&$  1.9\substack{+  1.5\\ -0.7}\times 10^{27}$\\
     48&SOL2020-02-17T08:34&10.2& 0.20&$ 5.5\substack{+ 0.5\\-0.5}$&$ 0.33\substack{+ 0.08\\-0.06}$& 0.123&           4&$  8.4\substack{+  1.3\\ -1.1}\times 10^{26}$\\
     49&SOL2020-02-23T02:38& 9.1& 0.16&$ 5.6\substack{+ 1.2\\-0.9}$&$ 0.11\substack{+ 0.06\\-0.04}$& 0.044&           8&$  8.4\substack{+  3.0\\ -2.0}\times 10^{26}$\\
     50&SOL2020-02-23T08:38& 9.4& 0.17&$ 6.2\substack{+ 0.7\\-0.7}$&$ 0.15\substack{+ 0.04\\-0.03}$& 0.076&          25&$  2.5\substack{+  0.5\\ -0.4}\times 10^{27}$\\
     51&SOL2020-02-23T19:54&18.5& 0.54&$ 6.9\substack{+ 0.7\\-0.7}$&$ 0.29\substack{+ 0.06\\-0.04}$& 0.191&           6&$  1.3\substack{+  0.2\\ -0.2}\times 10^{27}$\\
     52&SOL2020-02-23T22:12& 7.4& 0.10& - & - & - &           8 & - \\
     53&SOL2020-02-24T13:54& 7.7& 0.11& - & - & - &           8 & - \\
     54&SOL2020-02-27T17:14&10.2& 0.21&$ 5.3\substack{+ 0.6\\-0.6}$&$ 0.24\substack{+ 0.09\\-0.06}$&  0.081&- & - \\
     55&SOL2020-02-27T17:58&15.1& 0.32&$ 6.3\substack{+ 0.3\\-0.3}$&$ 0.37\substack{+ 0.04\\-0.03}$& 0.200&          17&$  3.0\substack{+  0.2\\ -0.2}\times 10^{27}$\\
     56&SOL2020-02-28T22:24&17.1& 0.39&$ 5.5\substack{+ 0.4\\-0.3}$&$ 0.43\substack{+ 0.07\\-0.07}$& 0.163&           7&$  1.5\substack{+  0.2\\ -0.1}\times 10^{27}$\\
     57&SOL2020-03-01T04:30& 8.1& 0.11&$ 6.7\substack{+ 0.9\\-0.8}$&$ 0.09\substack{+ 0.03\\-0.02}$& 0.055&          37&$  2.8\substack{+  0.6\\ -0.5}\times 10^{27}$\\
     58&SOL2020-03-02T05:56& 9.1& 0.16&$ 5.8\substack{+ 1.0\\-0.9}$&$ 0.15\substack{+ 0.07\\-0.04}$& 0.066&           5&$  7.1\substack{+  2.1\\ -1.5}\times 10^{26}$\\
     59&SOL2020-03-03T09:02& 9.2& 0.17&$ 5.5\substack{+ 0.3\\-0.3}$&$ 0.19\substack{+ 0.02\\-0.02}$&  0.071&- & - \\
     60&SOL2020-03-03T11:10& 7.2& 0.11&$ 2.8\substack{+ 1.0\\-0.6}$&$ 0.51\substack{+ 1.40\\-0.36}$& 0.013&          29&$  2.4\substack{+  3.4\\ -1.0}\times 10^{27}$\\
     61&SOL2020-03-04T09:56& 7.8& 0.15&$ 5.7\substack{+ 1.8\\-1.3}$&$ 0.11\substack{+ 0.10\\-0.04}$& 0.044&           4&$  5.0\substack{+  2.8\\ -1.5}\times 10^{26}$\\
     62&SOL2020-03-04T15:02& 7.1& 0.13&$ 5.5\substack{+ 7.2\\-1.4}$&$ 0.09\substack{+ 0.11\\-0.05}$& 0.032&           8&$  7.2\substack{+ 10.5\\ -2.9}\times 10^{26}$\\
     63&SOL2020-03-05T21:40&14.0& 0.35&$ 5.4\substack{+ 0.4\\-0.4}$&$ 0.39\substack{+ 0.07\\-0.06}$& 0.142&           6&$  1.2\substack{+  0.1\\ -0.1}\times 10^{27}$\\
     64&SOL2020-03-21T12:20&19.2& 0.47&$ 5.7\substack{+ 0.3\\-0.3}$&$ 0.82\substack{+ 0.11\\-0.09}$& 0.336&          38&$  7.4\substack{+  0.7\\ -0.6}\times 10^{27}$\\
     65&SOL2020-03-21T13:02&12.5& 0.23&$ 5.0\substack{+ 0.3\\-0.3}$&$ 0.49\substack{+ 0.08\\-0.07}$& 0.137&          40&$  5.2\substack{+  0.5\\ -0.5}\times 10^{27}$\\
     66&SOL2020-03-21T15:08&23.6& 0.61&$ 6.2\substack{+ 0.1\\-0.2}$&$ 0.90\substack{+ 0.06\\-0.04}$& 0.466&          31&$  7.3\substack{+  0.3\\ -0.3}\times 10^{27}$\\
     67&SOL2020-03-22T03:34& 7.5& 0.10&$ 4.6\substack{+ 1.9\\-1.7}$&$ 0.11\substack{+ 0.41\\-0.06}$& 0.022&           5&$  4.7\substack{+  9.4\\ -2.2}\times 10^{26}$\\
     68&SOL2020-03-22T04:18& 8.6& 0.17&$ 5.7\substack{+ 1.0\\-0.8}$&$ 0.18\substack{+ 0.08\\-0.05}$& 0.076&          28&$  2.8\substack{+  0.8\\ -0.5}\times 10^{27}$\\
     69&SOL2020-03-22T06:08& 8.0& 0.11&$ 5.5\substack{+ 1.1\\-1.0}$&$ 0.12\substack{+ 0.07\\-0.04}$& 0.043&          31&$  2.3\substack{+  0.9\\ -0.6}\times 10^{27}$\\
     70&SOL2020-03-23T10:57& 7.2& 0.11&$ 4.3\substack{+ 0.9\\-0.7}$&$ 0.20\substack{+ 0.18\\-0.09}$& 0.033&           6&$  7.0\substack{+  3.5\\ -1.9}\times 10^{26}$\\
     71&SOL2020-03-24T07:32& 6.9& 0.08&$ 6.0\substack{+ 3.4\\-2.6}$&$ 0.02\substack{+ 0.09\\-0.01}$&  0.010&- & - \\
     72&SOL2020-03-24T11:10& 7.9& 0.11&$ 5.5\substack{+ 1.6\\-1.3}$&$ 0.10\substack{+ 0.10\\-0.04}$&  0.035&- & - \\
     73&SOL2020-03-24T12:56& 9.8& 0.16&$ 4.7\substack{+ 0.6\\-0.6}$&$ 0.21\substack{+ 0.11\\-0.06}$& 0.049&           7&$  8.8\substack{+  2.5\\ -1.7}\times 10^{26}$\\
     74&SOL2020-03-24T14:24& 9.2& 0.14&$ 5.3\substack{+ 1.1\\-0.9}$&$ 0.13\substack{+ 0.08\\-0.04}$& 0.042&           5&$  5.9\substack{+  2.2\\ -1.4}\times 10^{26}$\\
     75&SOL2020-03-24T15:36& 7.8& 0.13&$ 2.8\substack{+ 1.0\\-0.6}$&$ 0.59\substack{+ 1.46\\-0.43}$&  0.014&- & - \\
     76&SOL2020-03-24T17:44& 7.7& 0.12&$ 3.7\substack{+ 0.9\\-0.7}$&$ 0.21\substack{+ 0.24\\-0.10}$& 0.020&           8&$  7.7\substack{+  4.7\\ -2.4}\times 10^{26}$\\
     77&SOL2020-03-24T22:02&10.2& 0.16&$ 6.4\substack{+ 0.7\\-0.6}$&$ 0.14\substack{+ 0.03\\-0.02}$& 0.075&          20&$  2.1\substack{+  0.3\\ -0.3}\times 10^{27}$\\
     78&SOL2020-03-25T01:56& 8.6& 0.14&$ 4.3\substack{+ 0.9\\-0.7}$&$ 0.21\substack{+ 0.17\\-0.09}$&  0.034&- & - \\
     79&SOL2020-03-25T05:28& 8.0& 0.13&$ 5.3\substack{+ 2.2\\-1.5}$&$ 0.05\substack{+ 0.09\\-0.03}$& 0.017&           4&$  3.2\substack{+  3.1\\ -1.3}\times 10^{26}$\\
     80&SOL2020-03-25T14:50& 9.0& 0.16&$ 5.2\substack{+ 1.2\\-0.9}$&$ 0.10\substack{+ 0.07\\-0.04}$& 0.034&           9&$  8.3\substack{+  3.4\\ -2.1}\times 10^{26}$\\
     81&SOL2020-03-25T17:14& 6.9& 0.11& - & - & - &          12 & - \\
     82&SOL2020-03-25T23:22& 8.4& 0.13&$ 3.9\substack{+ 2.1\\-1.2}$&$ 0.08\substack{+ 0.28\\-0.06}$& 0.009&           8&$  5.0\substack{+  8.9\\ -2.4}\times 10^{26}$\\
     83&SOL2020-03-26T05:00& 7.2& 0.12&$ 3.5\substack{+ 0.9\\-0.6}$&$ 0.34\substack{+ 0.42\\-0.19}$& 0.023&          25&$  2.1\substack{+  1.4\\ -0.7}\times 10^{27}$\\
     84&SOL2020-03-26T11:00&20.0& 0.57&$ 6.5\substack{+ 0.5\\-0.5}$&$ 0.61\substack{+ 0.10\\-0.07}$& 0.349&           5&$  1.6\substack{+  0.2\\ -0.2}\times 10^{27}$\\
     85&SOL2020-03-26T17:30& 8.7& 0.14&$ 5.8\substack{+ 0.9\\-0.7}$&$ 0.12\substack{+ 0.05\\-0.03}$& 0.050&           4&$  5.3\substack{+  1.3\\ -1.0}\times 10^{26}$\\
     86&SOL2020-03-28T04:18& 7.4& 0.13&$ 5.3\substack{+ 0.9\\-0.7}$&$ 0.14\substack{+ 0.07\\-0.04}$&  0.047&- & - \\
     87&SOL2020-04-13T05:28& 9.2& 0.16&$ 3.3\substack{+ 0.8\\-0.5}$&$ 0.43\substack{+ 0.51\\-0.23}$& 0.023&           6&$  7.7\substack{+  5.0\\ -2.5}\times 10^{26}$\\
     88&SOL2020-04-13T14:20& 8.5& 0.14& - & - & - &           7 & - \\
     89&SOL2020-04-15T05:20& 6.2& 0.08&$ 5.4\substack{+ 0.8\\-1.2}$&$ 0.06\substack{+ 0.06\\-0.02}$& 0.022&          24&$  1.4\substack{+  0.7\\ -0.4}\times 10^{27}$\\
     90&SOL2020-04-15T09:16& 7.4& 0.11&$ 4.6\substack{+ 0.7\\-0.6}$&$ 0.17\substack{+ 0.08\\-0.05}$& 0.038&          19&$  1.7\substack{+  0.5\\ -0.3}\times 10^{27}$\\
     91&SOL2020-04-15T20:44& 7.3& 0.15&$ 5.6\substack{+ 1.5\\-1.3}$&$ 0.10\substack{+ 0.10\\-0.04}$& 0.039&           6&$  6.3\substack{+  3.6\\ -1.9}\times 10^{26}$\\
     92&SOL2020-04-16T05:34& 8.6& 0.18&$ 6.3\substack{+ 0.7\\-0.6}$&$ 0.11\substack{+ 0.03\\-0.02}$& 0.061&           8&$  9.5\substack{+  1.6\\ -1.3}\times 10^{26}$\\
     93&SOL2020-04-17T14:02& 6.2& 0.08&$ 5.6\substack{+ 1.7\\-1.3}$&$ 0.09\substack{+ 0.08\\-0.04}$& 0.036&          10&$  9.0\substack{+  4.9\\ -2.7}\times 10^{26}$\\
     94&SOL2020-04-20T12:11& 6.8& 0.12&$ 5.3\substack{+ 1.4\\-1.0}$&$ 0.08\substack{+ 0.07\\-0.03}$& 0.029&           6&$  5.6\substack{+  2.6\\ -1.6}\times 10^{26}$\\
     95&SOL2020-04-21T05:10& 6.7& 0.08& - & - & - &          12 & - \\
     96&SOL2020-04-21T17:26& 7.9& 0.12&$ 4.4\substack{+ 0.7\\-0.5}$&$ 0.13\substack{+ 0.06\\-0.05}$&  0.025&- & - \\
     97&SOL2020-04-22T09:16&10.3& 0.21&$ 5.2\substack{+ 0.8\\-0.6}$&$ 0.18\substack{+ 0.08\\-0.05}$& 0.058&           5&$  7.0\substack{+  1.8\\ -1.3}\times 10^{26}$\\
     98&SOL2020-04-22T20:48& 7.3& 0.13&$ 5.9\substack{+ 1.4\\-1.1}$&$ 0.06\substack{+ 0.00\\-0.02}$& 0.025&          14&$  9.4\substack{+  2.3\\ -2.4}\times 10^{26}$\\

%% file: paper2_microflares_xsm.bbl
\begin{thebibliography}{}
\expandafter\ifx\csname natexlab\endcsname\relax\def\natexlab#1{#1}\fi
\providecommand{\url}[1]{\href{#1}{#1}}
\providecommand{\dodoi}[1]{doi:~\href{http://doi.org/#1}{\nolinkurl{#1}}}
\providecommand{\doeprint}[1]{\href{http://ascl.net/#1}{\nolinkurl{http://ascl.net/#1}}}
\providecommand{\doarXiv}[1]{\href{https://arxiv.org/abs/#1}{\nolinkurl{https://arxiv.org/abs/#1}}}

\bibitem[{{Arnaud}(1996)}]{arnaud96}
{Arnaud}, K.~A. 1996, in Astronomical Society of the Pacific Conference Series,
  Vol. 101, Astronomical Data Analysis Software and Systems V, ed. G.~H.
  {Jacoby} \& J.~{Barnes}, 17

\bibitem[{{Aschwanden} {et~al.}(2000{\natexlab{a}}){Aschwanden}, {Nightingale},
  {Tarbell}, \& {Wolfson}}]{2000ApJ...535.1027A}
{Aschwanden}, M.~J., {Nightingale}, R.~W., {Tarbell}, T.~D., \& {Wolfson},
  C.~J. 2000{\natexlab{a}}, \apj, 535, 1027, \dodoi{10.1086/308866}

\bibitem[{{Aschwanden} {et~al.}(2000{\natexlab{b}}){Aschwanden}, {Tarbell},
  {Nightingale}, {Schrijver}, {Title}, {Kankelborg}, {Martens}, \&
  {Warren}}]{2000ApJ...535.1047A}
{Aschwanden}, M.~J., {Tarbell}, T.~D., {Nightingale}, R.~W., {et~al.}
  2000{\natexlab{b}}, \apj, 535, 1047, \dodoi{10.1086/308867}

\bibitem[{{Benz}(2017)}]{2017LRSP...14....2B}
{Benz}, A.~O. 2017, Living Reviews in Solar Physics, 14, 2,
  \dodoi{10.1007/s41116-016-0004-3}

\bibitem[{{Benz} \& {Krucker}(2002)}]{2002ApJ...568..413B}
{Benz}, A.~O., \& {Krucker}, S. 2002, \apj, 568, 413, \dodoi{10.1086/338807}

\bibitem[{{Christe} {et~al.}(2008){Christe}, {Hannah}, {Krucker}, {McTiernan},
  \& {Lin}}]{2008ApJ...677.1385C}
{Christe}, S., {Hannah}, I.~G., {Krucker}, S., {McTiernan}, J., \& {Lin}, R.~P.
  2008, \apj, 677, 1385, \dodoi{10.1086/529011}

\bibitem[{{Cooper} {et~al.}(2020){Cooper}, {Hannah}, {Grefenstette},
  {Glesener}, {Krucker}, {Hudson}, {White}, \& {Smith}}]{2020ApJ...893L..40C}
{Cooper}, K., {Hannah}, I.~G., {Grefenstette}, B.~W., {et~al.} 2020, \apjl,
  893, L40, \dodoi{10.3847/2041-8213/ab873e}

\bibitem[{{Del Zanna}(2013)}]{2013A&A...558A..73D}
{Del Zanna}, G. 2013, \aap, 558, A73, \dodoi{10.1051/0004-6361/201321653}

\bibitem[{{Dere} {et~al.}(2019){Dere}, {Del Zanna}, {Young}, {Landi}, \&
  {Sutherland}}]{2019ApJS..241...22D}
{Dere}, K.~P., {Del Zanna}, G., {Young}, P.~R., {Landi}, E., \& {Sutherland},
  R.~S. 2019, \apjs, 241, 22, \dodoi{10.3847/1538-4365/ab05cf}

\bibitem[{{Dere} {et~al.}(1997){Dere}, {Landi}, {Mason}, {Monsignori Fossi}, \&
  {Young}}]{1997A&AS..125..149D}
{Dere}, K.~P., {Landi}, E., {Mason}, H.~E., {Monsignori Fossi}, B.~C., \&
  {Young}, P.~R. 1997, \aaps, 125, 149, \dodoi{10.1051/aas:1997368}

\bibitem[{{Glesener} {et~al.}(2017){Glesener}, {Krucker}, {Hannah}, {Hudson},
  {Grefenstette}, {White}, {Smith}, \& {Marsh}}]{2017ApJ...845..122G}
{Glesener}, L., {Krucker}, S., {Hannah}, I.~G., {et~al.} 2017, \apj, 845, 122,
  \dodoi{10.3847/1538-4357/aa80e9}

\bibitem[{{Glesener} {et~al.}(2020){Glesener}, {Krucker}, {Duncan}, {Hannah},
  {Grefenstette}, {Chen}, {Smith}, {White}, \& {Hudson}}]{2020ApJ...891L..34G}
{Glesener}, L., {Krucker}, S., {Duncan}, J., {et~al.} 2020, \apjl, 891, L34,
  \dodoi{10.3847/2041-8213/ab7341}

\bibitem[{{Golub} {et~al.}(1974){Golub}, {Krieger}, {Silk}, {Timothy}, \&
  {Vaiana}}]{1974ApJ...189L..93G}
{Golub}, L., {Krieger}, A.~S., {Silk}, J.~K., {Timothy}, A.~F., \& {Vaiana},
  G.~S. 1974, \apjl, 189, L93, \dodoi{10.1086/181472}

\bibitem[{{Golub} {et~al.}(2007){Golub}, {Deluca}, {Austin}, {Bookbinder},
  {Caldwell}, {Cheimets}, {Cirtain}, {Cosmo}, {Reid}, {Sette}, {Weber},
  {Sakao}, {Kano}, {Shibasaki}, {Hara}, {Tsuneta}, {Kumagai}, {Tamura},
  {Shimojo}, {McCracken}, {Carpenter}, {Haight}, {Siler}, {Wright}, {Tucker},
  {Rutledge}, {Barbera}, {Peres}, \& {Varisco}}]{2007SoPh..243...63G}
{Golub}, L., {Deluca}, E., {Austin}, G., {et~al.} 2007, \solphys, 243, 63,
  \dodoi{10.1007/s11207-007-0182-1}

\bibitem[{{Gryciuk} {et~al.}(2017){Gryciuk}, {Siarkowski}, {Sylwester},
  {Gburek}, {Podgorski}, {Kepa}, {Sylwester}, \&
  {Mrozek}}]{2017SoPh..292...77G}
{Gryciuk}, M., {Siarkowski}, M., {Sylwester}, J., {et~al.} 2017, \solphys, 292,
  77, \dodoi{10.1007/s11207-017-1101-8}

\bibitem[{{Hannah} {et~al.}(2008){Hannah}, {Christe}, {Krucker}, {Hurford},
  {Hudson}, \& {Lin}}]{2008ApJ...677..704H}
{Hannah}, I.~G., {Christe}, S., {Krucker}, S., {et~al.} 2008, \apj, 677, 704,
  \dodoi{10.1086/529012}

\bibitem[{{Hannah} {et~al.}(2011){Hannah}, {Hudson}, {Battaglia}, {Christe},
  {Ka{\v{s}}parov{\'a}}, {Krucker}, {Kundu}, \&
  {Veronig}}]{2011SSRv..159..263H}
{Hannah}, I.~G., {Hudson}, H.~S., {Battaglia}, M., {et~al.} 2011, \ssr, 159,
  263, \dodoi{10.1007/s11214-010-9705-4}

\bibitem[{{Hannah} {et~al.}(2019){Hannah}, {Kleint}, {Krucker}, {Grefenstette},
  {Glesener}, {Hudson}, {White}, \& {Smith}}]{2019ApJ...881..109H}
{Hannah}, I.~G., {Kleint}, L., {Krucker}, S., {et~al.} 2019, \apj, 881, 109,
  \dodoi{10.3847/1538-4357/ab2dfa}

\bibitem[{{Hudson}(1991)}]{1991SoPh..133..357H}
{Hudson}, H.~S. 1991, \solphys, 133, 357, \dodoi{10.1007/BF00149894}

\bibitem[{{Janardhan} {et~al.}(2011){Janardhan}, {Bisoi}, {Ananthakrishnan},
  {Tokumaru}, \& {Fujiki}}]{2011GeoRL..3820108J}
{Janardhan}, P., {Bisoi}, S.~K., {Ananthakrishnan}, S., {Tokumaru}, M., \&
  {Fujiki}, K. 2011, \grl, 38, L20108, \dodoi{10.1029/2011GL049227}

\bibitem[{{Janardhan} {et~al.}(2015){Janardhan}, {Bisoi}, {Ananthakrishnan},
  {Tokumaru}, {Fujiki}, {Jose}, \& {Sridharan}}]{2015JGRA..120.5306J}
{Janardhan}, P., {Bisoi}, S.~K., {Ananthakrishnan}, S., {et~al.} 2015, Journal
  of Geophysical Research (Space Physics), 120, 5306,
  \dodoi{10.1002/2015JA021123}

\bibitem[{{Klimchuk}(2006)}]{2006SoPh..234...41K}
{Klimchuk}, J.~A. 2006, \solphys, 234, 41, \dodoi{10.1007/s11207-006-0055-z}

\bibitem[{{Krucker} \& {Benz}(1998)}]{1998ApJ...501L.213K}
{Krucker}, S., \& {Benz}, A.~O. 1998, \apjl, 501, L213, \dodoi{10.1086/311474}

\bibitem[{{Krucker} {et~al.}(1997){Krucker}, {Benz}, {Bastian}, \&
  {Acton}}]{1997ApJ...488..499K}
{Krucker}, S., {Benz}, A.~O., {Bastian}, T.~S., \& {Acton}, L.~W. 1997, \apj,
  488, 499, \dodoi{10.1086/304686}

\bibitem[{{Krucker} {et~al.}(2020){Krucker}, {Hurford}, {Grimm}, {K{\"o}gl},
  {Gr{\"o}belbauer}, {Etesi}, {Casadei}, {Csillaghy}, {Benz}, {Arnold},
  {Molendini}, {Orleanski}, {Schori}, {Xiao}, {Kuhar}, {Hochmuth}, {Felix},
  {Schramka}, {Marcin}, {Kobler}, {Iseli}, {Dreier}, {Wiehl}, {Kleint},
  {Battaglia}, {Lastufka}, {Sathiapal}, {Lapadula}, {Bednarzik}, {Birrer},
  {Stutz}, {Wild}, {Marone}, {Skup}, {Cichocki}, {Ber}, {Rutkowski}, {Bujwan},
  {Juchnikowski}, {Winkler}, {Darmetko}, {Michalska}, {Seweryn}, {Bia{\l}ek},
  {Osica}, {Sylwester}, {Kowalinski}, {{\'S}cis{\l}owski}, {Siarkowski},
  {St{\k{e}}{\'s}licki}, {Mrozek}, {Podg{\'o}rski}, {Meuris}, {Limousin},
  {Gevin}, {Le Mer}, {Brun}, {Strugarek}, {Vilmer}, {Musset}, {Maksimovi{\'c}},
  {F{\'a}rn{\'\i}k}, {Koz{\'a}{\v{c}}ek}, {Ka{\v{s}}parov{\'a}}, {Mann},
  {{\"O}nel}, {Warmuth}, {Rendtel}, {Anderson}, {Bauer}, {Dionies}, {Paschke},
  {Pl{\"u}schke}, {Woche}, {Schuller}, {Veronig}, {Dickson}, {Gallagher},
  {Maloney}, {Bloomfield}, {Piana}, {Massone}, {Benvenuto}, {Massa},
  {Schwartz}, {Dennis}, {van Beek}, {Rodr{\'\i}guez-Pacheco}, \&
  {Lin}}]{2020A&A...642A..15K}
{Krucker}, S., {Hurford}, G.~J., {Grimm}, O., {et~al.} 2020, \aap, 642, A15,
  \dodoi{10.1051/0004-6361/201937362}

\bibitem[{{Kuhar} {et~al.}(2018){Kuhar}, {Krucker}, {Glesener}, {Hannah},
  {Grefenstette}, {Smith}, {Hudson}, \& {White}}]{2018ApJ...856L..32K}
{Kuhar}, M., {Krucker}, S., {Glesener}, L., {et~al.} 2018, \apjl, 856, L32,
  \dodoi{10.3847/2041-8213/aab889}

\bibitem[{{Leibacher} {et~al.}(2010){Leibacher}, {Sakurai}, {Schrijver}, \&
  {van Driel-Gesztelyi}}]{2010SoPh..263....1L}
{Leibacher}, J., {Sakurai}, T., {Schrijver}, C.~J., \& {van Driel-Gesztelyi},
  L. 2010, \solphys, 263, 1, \dodoi{10.1007/s11207-010-9553-0}

\bibitem[{{Lemen} {et~al.}(2012){Lemen}, {Title}, {Akin}, {Boerner}, {Chou},
  {Drake}, {Duncan}, {Edwards}, {Friedlaender}, {Heyman}, {Hurlburt}, {Katz},
  {Kushner}, {Levay}, {Lindgren}, {Mathur}, {McFeaters}, {Mitchell}, {Rehse},
  {Schrijver}, {Springer}, {Stern}, {Tarbell}, {Wuelser}, {Wolfson}, {Yanari},
  {Bookbinder}, {Cheimets}, {Caldwell}, {Deluca}, {Gates}, {Golub}, {Park},
  {Podgorski}, {Bush}, {Scherrer}, {Gummin}, {Smith}, {Auker}, {Jerram},
  {Pool}, {Soufli}, {Windt}, {Beardsley}, {Clapp}, {Lang}, \&
  {Waltham}}]{2012SoPh..275...17L}
{Lemen}, J.~R., {Title}, A.~M., {Akin}, D.~J., {et~al.} 2012, \solphys, 275,
  17, \dodoi{10.1007/s11207-011-9776-8}

\bibitem[{{Lin} {et~al.}(1984){Lin}, {Schwartz}, {Kane}, {Pelling}, \&
  {Hurley}}]{1984ApJ...283..421L}
{Lin}, R.~P., {Schwartz}, R.~A., {Kane}, S.~R., {Pelling}, R.~M., \& {Hurley},
  K.~C. 1984, \apj, 283, 421, \dodoi{10.1086/162321}

\bibitem[{{Madjarska}(2019)}]{2019LRSP...16....2M}
{Madjarska}, M.~S. 2019, Living Reviews in Solar Physics, 16, 2,
  \dodoi{10.1007/s41116-019-0018-8}

\bibitem[{{Mithun} {et~al.}(2020){Mithun}, {Vadawale}, {Sarkar}, {Shanmugam},
  {Patel}, {Mondal}, {Joshi}, {Janardhan}, {Adalja}, {Goyal}, {Ladiya},
  {Tiwari}, {Singh}, {Kumar}, {Tiwari}, {Modi}, \&
  {Bhardwaj}}]{2020SoPh..295..139M}
{Mithun}, N.~P.~S., {Vadawale}, S.~V., {Sarkar}, A., {et~al.} 2020, \solphys,
  295, 139, \dodoi{10.1007/s11207-020-01712-1}

\bibitem[{{Mithun} {et~al.}(2021){Mithun}, {Vadawale}, {Patel}, {Shanmugam},
  {Chakrabarty}, {Konar}, {Sarvaiya}, {Padia}, {Sarkar}, {Kumar}, {Jangid},
  {Sarda}, {Shah}, \& {Bhardwaj}}]{mithun20_soft}
{Mithun}, N.~P.~S., {Vadawale}, S.~V., {Patel}, A.~R., {et~al.} 2021, Astronomy
  and Computing, 34, 100449,
  \dodoi{https://doi.org/10.1016/j.ascom.2021.100449}

\bibitem[{{Parker}(1988)}]{1988ApJ...330..474P}
{Parker}, E.~N. 1988, \apj, 330, 474, \dodoi{10.1086/166485}

\bibitem[{{Parnell} \& {Jupp}(2000)}]{2000ApJ...529..554P}
{Parnell}, C.~E., \& {Jupp}, P.~E. 2000, \apj, 529, 554, \dodoi{10.1086/308271}

\bibitem[{{Priest} {et~al.}(1994){Priest}, {Parnell}, \&
  {Martin}}]{1994ApJ...427..459P}
{Priest}, E.~R., {Parnell}, C.~E., \& {Martin}, S.~F. 1994, \apj, 427, 459,
  \dodoi{10.1086/174157}

\bibitem[{{Sakurai}(2017)}]{2017PJAB...93...87S}
{Sakurai}, T. 2017, Proceeding of the Japan Academy, Series B, 93, 87,
  \dodoi{10.2183/pjab.93.006}

\bibitem[{{Scherrer} {et~al.}(2012){Scherrer}, {Schou}, {Bush}, {Kosovichev},
  {Bogart}, {Hoeksema}, {Liu}, {Duvall}, {Zhao}, {Title}, {Schrijver},
  {Tarbell}, \& {Tomczyk}}]{2012SoPh..275..207S}
{Scherrer}, P.~H., {Schou}, J., {Bush}, R.~I., {et~al.} 2012, \solphys, 275,
  207, \dodoi{10.1007/s11207-011-9834-2}

\bibitem[{{Shanmugam} {et~al.}(2020){Shanmugam}, {Vadawale}, {Patel},
  {Adalaja}, {Mithun}, {Ladiya}, {Goyal}, {Tiwari}, {Singh}, {Kumar},
  {Painkra}, {Acharya}, {Bhardwaj}, {Hait}, {Patinge}, {Kapoor}, {Kumar},
  {Satya}, {Saxena}, \& {Arvind}}]{shanmugam20}
{Shanmugam}, M., {Vadawale}, S.~V., {Patel}, A.~R., {et~al.} 2020, Current
  Science, 118, 45, \dodoi{10.18520/cs/v118/i1/45-52}

\bibitem[{{Shimizu}(1995)}]{shimizu_95}
{Shimizu}, T. 1995, \pasj, 47, 251

\bibitem[{{Sylwester} {et~al.}(2019){Sylwester}, {Sylwester}, {Siarkowski},
  {Phillips}, {Podgorski}, \& {Gryciuk}}]{2019SoPh..294..176S}
{Sylwester}, B., {Sylwester}, J., {Siarkowski}, M., {et~al.} 2019, \solphys,
  294, 176, \dodoi{10.1007/s11207-019-1565-9}

\bibitem[{{Vadawale} {et~al.}(2014){Vadawale}, {Shanmugam}, {Acharya}, {Patel},
  {Goyal}, {Shah}, {Hait}, {Patinge}, \& {Subrahmanyam}}]{2014AdSpR..54.2021V}
{Vadawale}, S.~V., {Shanmugam}, M., {Acharya}, Y.~B., {et~al.} 2014, Advances
  in Space Research, 54, 2021, \dodoi{10.1016/j.asr.2013.06.002}

\bibitem[{{Wright} {et~al.}(2017){Wright}, {Hannah}, {Grefenstette},
  {Glesener}, {Krucker}, {Hudson}, {Smith}, {Marsh}, {White}, \&
  {Kuhar}}]{wright_17}
{Wright}, P.~J., {Hannah}, I.~G., {Grefenstette}, B.~W., {et~al.} 2017, \apj,
  844, 132, \dodoi{10.3847/1538-4357/aa7a59}

\end{thebibliography}
